\documentstyle[aps,multicol,epsfig]{revtex}
\draft
\begin{document}
\title{Kondo effect in real quantum dots}
\author{M. Pustilnik and L. I. Glazman \vspace{1.2mm}}
\address{Theoretical Physics Institute, University of Minnesota, 
Minneapolis, MN 55455 \vspace{1.2mm}}
\maketitle

\begin{abstract}
Exchange interaction within a quantum dot strongly affects the
transport through it in the Kondo regime. In a striking difference
with the results of the conventional model, where this interaction is
neglected, here the temperature and magnetic field dependence of the
conductance may become non-monotonic: its initial increase follows by
a drop when temperature and magnetic field are lowered.  
\end{abstract}

\pacs{PACS numbers: 72.15.Qm, 73.23.-b, 73.23.Hk, 73.63.Kv}

\vspace{-3mm}
\begin{multicols}{2}

The conventional theory of the Kondo effect in
tunneling~\cite{classics} is based on the Anderson impurity
model~\cite{Anderson}. The use of this model for describing the
electron transport through a conducting grain was first suggested
in~\cite{GZ}, and later successfully applied in the context of quantum
dots in the Coulomb blockade regime (see, {\it e.g.},~\cite{MWL}). In
these applications, the quantum dot is modeled by one singly occupied
energy level. The model predicts monotonic increase of the tunneling
conductance with lowering the temperature. By now, a strong
experimental evidence exists for such behavior~\cite{exp}.

To measure transport across the dot, at least two leads should be
attached to it. The problem of transmission through a dot is similar
to that of transition between channels in a multichannel scattering
problem. If the system is described by the Anderson model, then the
corresponding transition rate turns out to be proportional to the
total scattering cross-section. Therefore the two problems,
calculation of the tunneling conductance $G$, and evaluation of the
magnetic impurity contribution to the resistivity of a bulk
metal, are equivalent~\cite{AM}. Well below the Kondo
temperature, the scattering cross-section reaches the unitary
limit; accordingly, $G$ saturates at the level
$G\sim e^2/\pi\hbar$.

In a real quantum dot, the truncation of the spectrum to a single
level is not always possible. For example, dots with ground state spin
$S>1/2$ are not described by the Anderson impurity model\cite{weis}.
In this paper we demonstrate that in general the problem of transport
through a dot does not map onto the problem of resistivity in a bulk
metal. In the most interesting case, the tunneling conductance first
rises and than drops when temperature is lowered. The conductance
dependence on the magnetic field is also non-monotonic.

The confining potential forming a lateral quantum dot is smooth. The
dot-lead junction is essentially an electron waveguide\cite{ABG}.
Making it narrower, one pinches the propagating modes off. Coulomb
blockade develops when the last propagating mode is near its
pinch-off.  Therefore, for lateral dots in the Coulomb blockade regime
the number of channels per junction is one.

We start with a discussion of the dependence of the zero-temperature
conductance on the magnetic field applied in the plane of the dot. The
in-plane field results in the Zeeman splitting $B$ of the spin states
of the dot, but barely affects the orbital degrees of freedom. At a
finite $B$, the ground state of the system is not degenerate;
therefore, at $T=0$ an electron experiences only potential scattering.
The amplitudes of scattering ${\cal S}_{s;\alpha \alpha'}$ of
electrons with spin $s$ from lead $\alpha$ to lead $\alpha'$ form the
scattering matrix (here $\alpha,\alpha'=R\mbox{ or }L$ for the right
or left lead, respectively). It can be diagonalized by rotation in the
$R-L$ space to the new basis of channels $a$ and $b$,
\begin{equation}
U {\cal S}_s U^\dagger 
= {\rm diag}\left\{e^{2i\delta_{\gamma s}}\right\},
\quad
\gamma=a,b, \quad s=\pm 1.
\label{phases}
\end{equation}
Here $U =\exp(i\vartheta \tau^y)\exp(i\varphi \tau^z)$, and $\tau^i$
are the Pauli matrices acting in the $R-L$ space
($\tau^+=\tau^x+i\tau^y$ transforms $L\to R$). The parameters
$\vartheta$ and $\varphi$ depend on the microscopic properties of the
dot-lead junctions. The $T=0$ conductance is given by the Landauer
formula,
\begin{equation}
G(T=0) = \frac{e^2}{2\pi \hbar}\sum_s \left|{\cal S}_{s; RL}^2\right|,
\label{Landauer}
\end{equation}
and can be expressed now in terms of the scattering phase shifts 
$\delta_{\gamma s}$ at the Fermi level:
\begin{equation}
\frac{G}{G_0} = \frac{1}{2} \sum_s \sin^2(\delta_{as} - \delta_{bs}),
\quad
G_0 = \frac{e^2}{\pi \hbar}\sin^2(2\vartheta). 
\label{zeroT}
\end{equation}
(note that the conductance is independent of $\varphi$). 

In order to calculate the phase shifts, one needs to know the
effective Hamiltonian ${\cal H}$ of the system. Clearly, the term in
it representing the interaction of itinerant electrons with the spin
of the dot should obey the SU(2) symmetry,
\begin{equation}
{\cal H} = \sum_{\gamma ks}\xi_k \psi_{\gamma ks}^\dagger \psi_{\gamma ks} 
+  \sum_\gamma J_\gamma ({\bf s}_\gamma \cdot {\bf S}) - B S^z, 
\label{kondo}
\end{equation}
where ${\bf s}_\gamma = \sum_{kk'ss'}\psi_{\gamma ks}^{\dagger }
({{\bbox \sigma }_{ss'}}/{2}) \psi_{\gamma k's'}$ are local spin
densities of itinerant electrons. The last term in ${\cal H}$
describes the effect of the magnetic field. The Hamiltonian
(\ref{kondo}) does not include the potential scattering term; we defer
the discussion of its role till later in the paper. The description by
Eq.~(\ref{kondo}) is valid at energies below the energy gap $\Delta$
for spin excitations in the dot, which is of the order of the
single-particle level spacing in it. Note that the symmetry allows for
any signs of the exchange constants $J_{\gamma}$. We will present a
microscopic derivation of Eq.~(\ref{kondo}) towards the end of the
paper.

Following \cite{N}, we remove the ambiguity in the definition of the
phase shifts by fixing $\delta_{\gamma s}=0$ at $J_\gamma = 0$; then
$|\delta_{\gamma s}|\leq \pi/2$ at any $J_\gamma$. Furthermore, the
invariance of the Hamiltonian~(\ref{kondo}) with respect to the
particle-hole transformation, 
$\psi_{\gamma ks}\to s\psi_{\gamma -k-s}^\dagger$, 
yields the relation ${\cal S}_{s} {\cal S}_{-s}  = 1$ for the scattering 
matrix, which 
allows us to represent the phase shifts as
\begin{equation}
\delta_{\gamma s} = -s \delta_\gamma,\quad 0\leq\delta_\gamma\leq\pi/2.
\label{new_delta}
\end{equation}
Substitution into Eq.~(\ref{zeroT}) then gives 
\begin{equation}
G/G_0 = \sin^2(\delta_a - \delta_b).
\label{PH}
\end{equation}

In order to find $\delta_{\gamma}$ at $B= 0$, we
need to invoke the properties of the ground state of the Kondo model. 
We start with $S>1/2$ on the dot. In the case of an antiferromagnetic 
exchange in a channel, $J_\gamma>0$, the itinerant electrons
participate in the screening of the localized spin, thus reducing its 
value by $1/2$. By the Friedel sum rule, the corresponding phase shift
$\delta_\gamma=\pi/2$. The channels with the ferromagnetic coupling to
the localized spin, $J_\gamma<0$, decouple at low energies, and 
$\delta_\gamma=0$. It is then clear from Eq.~(\ref{PH}) that for $S>1/2$
the conductance at $T=0$ and $B=0$ is enhanced due to the Kondo effect
only if the two exchange constants have opposite signs.

At $S=1/2$ the enhancement of the conductance occurs also if both
exchange constants are positive. Indeed, the local spin is screened by
the channel, say $a$, with the larger exchange constant, so that
$\delta_a=\pi/2$. The other channel decouples~\cite{NB} at low
energies, and $\delta_b=0$. Note that in the special case $J_a=J_b$ the
conductance across the dot is zero, which is obvious from the
rotational symmetry in the $R-L$ space, existing in this case.

There is no mapping between the Kondo problems for tunneling and for
scattering in the bulk, except if one of the exchange constants in
Eq.~(\ref{kondo}) is zero. However, if the constants are finite and
have opposite signs (or both are positive at $S=1/2$), the conductance
behavior is qualitatively similar to the conventional~\cite{AM} one:
$G(T)$ and $G(B)$ increase monotonously when $T$ and $B$ are lowered,
eventually saturating at the value $G\sim e^2/\pi\hbar$. If both
coupling constants are negative (fully ferromagnetic exchange
interaction) there is no Kondo effect: $G$ decreases when $T$ and $B$
are lowered.

We concentrate now on the most interesting case of the
antiferromagnetic coupling, $J_a\geq J_b>0$, and spin $S\geq 1$.  In
this case, the Kondo effect manifests itself in a peculiar way: the
field and temperature dependence of $G$ is not monotonic, which
differs qualitatively form the conventional picture. Here we present a
detailed analysis only of the dependence $G(B)$ at $T=0$.

\begin{figure}[tbp]
\centerline{\epsfxsize=5.5cm
\epsfbox{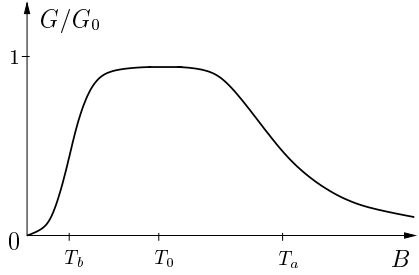}\vspace{2.5mm}}
\caption{ 
  Magnetic field dependence of the Kondo contribution to the
  conductance through a lateral quantum dot with $S>1/2$. The
  temperature dependence of the conductance exhibits a similar
  behavior.  }
\label{suppress}
\end{figure}

Finite field leads to deviations from the unitary limit
$\delta_{a,b}=\pi/2$. The magnitude of the deviations is controlled by
two parameters, $B/T_a$ and $B/T_b$, where $T_a\geq T_b$ are the Kondo
temperatures corresponding to the two exchange constants. For $S=1$
and $B = 0$ the dot's spin is screened completely.  If the Zeeman
energy is small, $B\ll T_b$, then the phase shifts are calculated by a
straightforward application of the Fermi-liquid technique\cite{N,NB},
and we find from Eq.~(\ref{PH})
\begin{equation}
\frac{G}{G_0} = \left(\frac{B}{T_a} -\frac{B}{T_b}\right)^2.
\label{FL}
\end{equation}
For $S>1$ the screening is incomplete: at the fixed point
$\delta_{a,b}=\pi/2$, and the remaining localized spin is $(S-1)$. The
fixed point is reached at $B=0$, and the approach to it is governed by
the ferromagnetic interaction of the remaining spin with itinerant
electrons\cite{NB}. The corresponding corrections can be calculated
with the logarithmic accuracy\cite{PWA}, yielding the asymptote of the
conductance in the regime $B\ll T_b$,
\begin{equation}
\frac{G}{G_0} =
\frac{\pi^2}{4}(S-1)^2 \left[\frac{1}{\ln(B/T_a)} - \frac{1}{\ln(B/T_b)}\right]^2.
\label{strong}
\end{equation}
In the weak coupling regime, $B\gg T_a$, one readily obtains for
arbitrary $S\neq 0$ (also with logarithmic accuracy)
\begin{equation}
\frac{G}{G_0} =
\frac{\pi^2}{4}S^2 \left[\frac{1}{\ln(B/T_a)} - \frac{1}{\ln(B/T_b)}\right]^2.
\label{weak}
\end{equation}
The asymptotes (\ref{FL}), (\ref{strong}) and (\ref{weak}) clearly
demonstrate that the dependence of $G(B)$ is non-monotonous with a maximum
at $B=T_0$ where $T_b\leq T_0\leq T_a$. This maximum can be
studied in detail if $T_b\ll T_a$. In this case, the intermediate
($T_b\ll B\ll T_a$) asymptote for $G(B)$ is
\begin{equation}
\frac{G}{G_0} = 1 - 
\frac{\pi^2}{4}(S-1/2)^2 \left[\frac{1}{\ln(B/T_a)} - 
\frac{1}{\ln(B/T_b)}\right]^2,
\label{mixed}
\end{equation}
and it displays maximum at $T_0 = \sqrt{T_aT_b}$. The conductance is
expected to be featureless in the crossover regions ($B\sim T_a$ or
$B\sim T_b$). The dependence of the conductance on the magnetic field
is shown schematically in Fig.~\ref{suppress}. It can be cast into the
form
\begin{equation}
G=G_0F_S\left(B/T_a, B/T_b\right).
\label{FS}
\end{equation}
The asymptotes of $F_S$ are given in Eqs.~(\ref{FL})-(\ref{mixed}).

The dependence of the conductance on temperature $G(T)$ is very
similar to $G(B)$. The asymptotes for $G(T)$ in the leading
logarithmic approximation can be obtained from
Eqs.~(\ref{strong})-(\ref{mixed}) by replacing $B\to T$, and, in
addition to it, $(S-1)^2\to S(S-1)$, $S^2\to S(S+1)$, and
$(S-1/2)^2\to S^2-1/4$ in Eqs.~(\ref{strong}), (\ref{weak}), and
(\ref{mixed}), respectively. The asymptote for $G(T)$ in the Fermi
liquid regime ($S=1,~T\ll T_b$) follows from Eq.~(\ref{FL}) after the
substitution $B\to\pi T$. The derivation of these asymptotes is
straightforward, but rather lengthy \cite{PG}, since the conductance
at a finite temperature is not expressed via elastic scattering
phase shifts [cf. Eq.~(\ref{zeroT})].

For the model described by Eq.~(\ref{kondo}) conductance $G$ vanishes 
in the limit $J_\gamma\to 0$. Yet, it is well-known that $G$ may remain 
finite even in this limit due to the cotunneling processes, not involving
the spin degrees of freedom~\cite{ABG}. These processes are described
by adding to the Hamiltonian~(\ref{kondo}) a potential scattering 
term $H_V$. The combination of the Kondo and potential scattering
yields
\begin{equation}
G=\widetilde{G}_0 F_S\left(B/\widetilde T_a,B/\widetilde T_b\right)+G_{el},
\label{FSct}
\end{equation}
instead of Eq.~(\ref{FS}). Here we sketch the derivation of
Eq.~(\ref{FSct}) for a simplified model in which $H_V$ is diagonal in
the $a-b$ basis,
\begin{equation}
H_V = \sum_\gamma V_\gamma\sum_{kk's} \psi_{\gamma ks}^\dagger\psi_{\gamma k's}
\label{potential}
\end{equation}
(the general case will be discussed elsewhere \cite{PG}). Potential
scattering results in finite phase shifts, $\delta_{\gamma
  s}=\delta_\gamma^0$, at $J_\gamma =0$; the corresponding elastic
cotunneling conductance is $G_{el} =
G_0\sin^2(\delta^0_a-\delta^0_b)$, see Eq.~(\ref{zeroT}).  In order to
consider now $J_\gamma \neq 0$, one can rewrite the Hamiltonian ${\cal
  H}+H_V$ in the new basis\cite{N}, which accounts for the phase
shifts $\delta^0_\gamma$. In this basis, the Hamiltonian again assumes
the form of Eq.~(\ref{kondo}), but with the modified\cite{N} exchange
constants $\widetilde J_{\gamma}= J_{\gamma}\cos^2\delta_\gamma^0$, 
which define new Kondo temperatures $\widetilde T_\gamma$. 
The phase shifts with respect to the new basis are $\tilde
\delta_{\gamma s}= -s \tilde \delta_\gamma$ [cf.
Eq.~(\ref{new_delta})]. The phase shifts $\tilde\delta_\gamma$ depend
on the magnetic field $B$ via two parameters, $B/\widetilde T_a$ and
$B/\widetilde T_b$.  Returning to the original basis, one finds~\cite{N} for
the phase shifts $\delta_{\gamma s} = \delta_{\gamma}^0 -
s\tilde\delta_{\gamma}$. Substitution of $\delta_{\gamma s}$ into
Eq.~(\ref{zeroT}) then yields Eq.~(\ref{FSct}) with $\widetilde{G}_0 =
G_0 - 2G_{el}$. Microscopic calculation~\cite{PG} performed for two
specific models, the Anderson model and the model of almost open
dot-lead junctions\cite{FurMat}, gives $\widetilde{G}_0 \geq 0$.

So far, we assumed that both exchange constants $J_\gamma$ in the
Hamiltonian (\ref{kondo}) differ from zero. Such an assumption was
made in a number of papers, see, {\it e.g.}, Ref.~\cite{Kroha}.
However, if a quantum dot is modelled by the Anderson
impurity~\cite{Anderson}, then the corresponding effective Hamiltonian
(\ref{kondo}) inevitably has only one non-zero exchange
constant~\cite{AM,ABG}.  We will demonstrate now that in fact both
exchange constants are finite, if Eq.~(\ref{kondo}) is derived
directly from a generic model of a quantum dot. The microscopic
Hamiltonian of the system consists of the part describing the free
electrons in the leads:
\begin{equation}
H_l =  \sum_{\alpha ks}\xi_{k} c_{\alpha ks}^\dagger c_{\alpha ks},
\quad
\alpha= R,L, 
\label{Hl}
\end{equation}
the tunneling part:
\begin{equation}
H_t = 
\sum_{\alpha nks}\left( t_{\alpha n}c_{\alpha ks}^\dagger d_{ns} + 
t_{\alpha n}^*d_{ns}^\dagger c_{\alpha ks}\right) , 
\label{Ht}
\end{equation}
and the Hamiltonian of the dot:
\begin{equation}
H_d =\sum_{ns}\epsilon _{n}d_{ns}^{\dagger }d_{ns} + H_{int}.
\label{Hd}
\end{equation}
The first term in the r.h.s. of Eq.~(\ref{Hd}) is the single-particle
part of $H_d$. The second term, $H_{int}$, represents the
electron-electron interaction within the dot. 
Extensive numerical renormalization group calculations
based on such a model were performed in Ref.~\cite{izumida}, where 
a non-monotonic temperature dependence
of the conductance was found for a special choice of $H_{int}$ and $t_{\alpha n}$.
However, the study~\cite{izumida} did not reveal 
the two-stage character of the Kondo screening, which shows that the 
non-monotonic behavior is a general feature, rather then a result of model assumptions.

Hamiltonian $H_d$ commutes with the total number of electrons in the 
dot, $\hat{N} = \sum_{ns}d_{ns}^{\dagger}d_{ns}$, and with its total 
spin, 
${\hat {\bf S}_{\rm tot}}
= \sum_{nss'}d_{ns}^{\dagger}({\bbox \sigma}_{ss'}/2)d_{ns'}$ 
(we neglect the weak spin-orbit interaction here).  The ground state of $H_d$ is 
characterized by the number of electrons $N$ and by spin $S$. In the absence 
of the magnetic field, the ground state is $(2S+1)$--fold degenerate and 
SU(2)--symmetric. This can be represented by introduction of the 
operator of spin of the dot in the ground state [cf. Eq.~(\ref{kondo})],
\begin{equation}
{\bf S}={\cal P}\hat{\bf S}_{\rm tot}{\cal P},
\label{SPIN} 
\end{equation}
where ${\cal P}$ is the projector onto the ground state manifold.

Tunneling Hamiltonian $H_t$ mixes the ground state with the states
having $N\pm 1$ electrons on the dot. For brevity we assume that the
dot is tuned to the middle of the Coulomb blockade valley, {\it i.e.},
its energy is increasing by the same amount $E_C$ if the number of
electrons is changed by $\pm 1$.  The energy deficit $E\sim E_C$ for
the corresponding transitions is high, which allows us to apply the
second-order perturbation theory in $H_t$ (Schrieffer-Wolff
transformation). The resulting effective Hamiltonian, which is valid at
energies less than the gap for intra-dot spin excitations $\Delta\ll
E_C$, has the form
\begin{equation}
H = H_l + \frac{4}{E_C}\sum_{\alpha \alpha' mn}t_{\alpha m}^* t_{\alpha' n} 
({\bf s}_{\alpha' \alpha} \cdot {\cal P}\hat{\bf S}_{mn}{\cal P}),  
\label{SW1}
\end{equation}
where
\[
{\bf s}_{\alpha \alpha'} = \sum_{kk'ss'}c_{\alpha ks}^{\dagger } 
\frac{{\bbox \sigma }_{ss'}}{2} c_{\alpha' k's'},
\quad
\hat{\bf S}_{mn} = \sum_{ss'}
d_{ms}^{\dagger }\frac{{\bbox \sigma }_{ss'}}{2}d_{ns'}.
\]
In the derivation of (\ref{SW1}) we have neglected the potential
scattering terms associated with the elastic cotunneling.

By SU(2) symmetry, the operators ${\cal P}\hat{\bf S}_{mn}{\cal P}$
for any $m$ and $n$ should be proportional to ${\bf S}$ given by
Eq.~(\ref{SPIN}):
\begin{equation}
{\cal P}\hat{\bf S}_{mn}{\cal P}= \Lambda_{mn}{\bf S}.
\label{Lambda}
\end{equation}
This allows us to rewrite Eq.~(\ref{SW1}) as
\begin{equation}
H= H_l + \sum_{\alpha \alpha'} J_{\alpha\alpha'}
({\bf s}_{\alpha'\alpha} \cdot {\bf S}), 
\label{SW2} 
\end{equation}
where the matrix of the exchange constants is
\begin{equation}
J_{\alpha\alpha'} = \frac{4}{E_C}
\sum_{mn}t_{\alpha m}^*t_{\alpha' n}\Lambda_{mn}.
\label{matrix}
\end{equation}
Note that the factors $t_{\alpha n}$ here depend on the properties of
the tunneling junctions, whereas $\Lambda_{mn}$ characterize the ground
state of the isolated dot, and depend on the electron-electron
interaction term $H_{int}$ in Eq.~(\ref{Hd}).

We can diagonalize matrix $J_{\alpha\alpha'}$, and thereby bring the
Hamiltonian (\ref{SW2}) to the form of Eq.~(\ref{kondo}). To do that,
we perform the unitary transformation $\psi_{\gamma ks} = \sum_\alpha
U_{\gamma\alpha}c_{\alpha ks}$, where $U_{\gamma\alpha}$ is the same
unitary matrix as in Eq.~(\ref{phases}). The eigenvalues of the matrix
$J_{\alpha\alpha'}$ are $J_a$ and $J_b$ of Eq.~(\ref{kondo}).

The matrix $\Lambda_{mn}$ of Eq.~(\ref{Lambda}) obviously satisfies
the relations $\Lambda =\Lambda^\dagger$ and ${\rm Tr}\Lambda = 1$. For a
generic choice of the tunneling amplitudes $t_{\alpha n}$ these
conditions alone fix neither the signs nor the magnitudes of the
exchange amplitudes $J_\gamma$. To make further progress one has to
rely upon assumptions regarding the form of $H_{int}$ in
(\ref{Hd}).  Although model realizations yielding any desired sign of
$J_\gamma$ can be easily constructed\cite{PG}, here we consider only
the simplest example,
\begin{equation}
H_{int} = E_C (\hat {N} - {\cal N})^2 - E_S \hat {\bf S}_{\rm tot}^2,
\label{RMT}
\end{equation}
which nevertheless results in the most interesting case $J_\gamma >0$,
$S\geq 1$. The parameter $E_S$ characterizes the strength of the
intradot exchange interaction; $E_S$ is of the order of the
single-particle level spacing in the dot. In the middle of the Coulomb
blockade valley, the dimensionless gate voltage ${\cal N}$ is an
integer, so the dot in the ground state has $N = {\cal N}$
electrons. The model~(\ref{RMT}) is fully justified\cite{ABG,spin}
only for large dots with chaotic single-particle states; the relative
magnitude of terms which are not included in
Eq.~(\ref{RMT}), is $\sim N^{-1/4}\ll 1$ (we assume here ballistic
motion of electrons within the dot).

For the Hamiltonian (\ref{Hd}), (\ref{RMT}), occupations of the
single-particle energy levels are good quantum numbers. In the ground
state with spin $S$, there are $2S$ singly occupied energy levels,
while all other levels are either empty or are occupied by a pair of
electrons with opposite spins. The matrix $\Lambda$ is diagonal, and
its only non-zero elements are the ones corresponding to the singly
occupied levels, $\Lambda_{mn} = (1/2S) \delta_{mn}$. Using
Eq.~(\ref{matrix}), it is easy to calculate the determinant of the
matrix $J_{\alpha\alpha'}$:
\begin{equation}
J_aJ_b = \frac{2}{E_C^2 S^2} \sum_{m,n=1}^{2S}
\left|(t_{Lm}t_{Rn}-t_{Rm}t_{Ln})^2\right| .
\label{det}
\end{equation}
For $S>1/2$ the sum in the r.h.s. here contains terms with $m\neq n$.
For a generic choice of tunneling amplitudes, these terms are
positive, and $J_aJ_b> 0$. The trace of
$J_{\alpha\alpha'}$, 
\begin{equation}
J_a+J_b = \frac{2}{E_C S} \sum_{\alpha}\sum_{m=1}^{2S}
\left|t_{\alpha m}^2\right|,
\label{trace}
\end{equation}
is also positive. Therefore, eigenvalues $J_{a,b}$ are positive, and
at $S>1/2$ both channels are coupled to the dot via antiferromagnetic
exchange interaction.

In the case of model interaction (\ref{RMT}) and the dot spin $S=1/2$,
only one discrete level is singly occulied. Then Eqs.~(\ref{det}) and
(\ref{trace}) yield $J_a> 0$ and $J_b=0$, just like in the Anderson
model. Accounting for corrections~\cite{ABG} to the Hamiltonian
(\ref{RMT}) results in finite values of both constants, $|J_b|\sim
J_a N^{-1/4}$.

The two characteristic temperatures, $T_a$ and $T_b$, should be of the
same order to allow for an observation of the predicted here
non-monotonic temperature dependence $G(T)$. Because of the
exponential dependence of $T_{a,b}$ on the corresponding exchange
constants, the two temperatures are very sensitive to the values of
the tunneling amplitudes for specific discrete levels, see
Eqs.~(\ref{matrix}), (\ref{det}), and (\ref{trace}); these values
hardly can be predicted in advance. However, one can tune the
amplitudes by applying a weak magnetic field {\it perpendicular} to
the plane of the dot. (It is sufficient to thread about one flux
quantum through the area of the dot\cite{Marcus}; in the case of GaAs,
such a field has a negligible Zeeman effect.) The continuous tuning
should allow one to go from a conventional manifestation~\cite{weis}
of the Kondo effect at $T_b\ll T_a$, to a non-monotonic $G(T)$ at
$T_b\lesssim T_a$, to a total suppression of the Kondo contribution to
the conductance at some value of the field, where $T_b=T_a$~\cite{induced}. 


To conclude, in this paper we considered electron transport through
lateral quantum dots. We predict that if the spin of the dot $S>1/2$,
then the dependence of conductance on temperature and in-plane
magnetic field is non-monotonic. 
This work was supported by NSF Grant DMR-9731756.

\end{multicols}

\begin{references}
\bibitem{induced} Tunneling amplitudes can be controlled without 
the use of a magnetic field in quantum dot systems of 
special geometry, such as two quantum dots connected in series.
\end{references}
\end{document}